\documentclass[mathleft,fleqn,%
]{an1}
\usepackage{graphicx}
\usepackage[varg]{txfonts}
\def \lsi {LSI+61$^{\circ}$303}

\def \rsun {$R_\odot$}
\def \msun {$M_\odot$}

\begin{document}

\Pagespan{1}{}
\Yearpublication{2017}%
\Yearsubmission{2016}%
\Month{1}%
\Volume{339}%
\Issue{1}%
\DOI{asna.201700000}%

\title{Mass of the compact object in the Be/gamma-ray binaries \lsi\ and MWC~148}

\author{R.\,K. Zamanov \inst{1}\fnmsep\thanks{Corresponding author:
        {rkz@astro.bas.bg}}
\and  J. Mart\'{\i}\inst{2}
\and  M. T. Garc\'{\i}a-Hern\'andez\inst{2}
}
\titlerunning{Compact object in \lsi\ and MWC~148}
\authorrunning{Zamanov, Mart\'{\i} \&  Garc\'{\i}a-Hern\'andez}
\institute{ Institute of Astronomy and National Astronomical Observatory, Bulgarian Academy of Sciences, Tsarigradsko Shose 72, 
BG-1784 Sofia, Bulgaria
\and 
Departamento de F\'isica, Escuela Polit\'ecnica Superior de Ja\'en, Universidad de Ja\'en, 
Campus Las Lagunillas,  A3, 23071, Ja\'en, Spain}

\received{27 September 2016}
\accepted{...}
\publonline{...}

\keywords{Stars: emission-line, Be -- binaries: spectroscopic -- Gamma rays: stars -- 
          Stars: individual: LSI+61303,  MWC~148}

\abstract{
We estimate the mass of the compact object in the  $\gamma$-ray binaries \lsi\ and MWC~148, using 
the latest data for the inclination, orbital motion and 
assuming that the orbital plane coincides with the equatorial plane of the Be star. 
For \lsi\ we find the mass of the compact object to be most likely in the range $1.3$~\msun $ < M_2 < 2.0$~\msun, 
which means that it is probably a neutron star.
For MWC~148, we find the mass of the compact object in a higher range, $2.1$~\msun $ < M_2 < 7.3$~\msun, 
which increases the chances for this system to host a black hole companion.}

\maketitle

\section{Introduction}
The $\gamma$-ray binaries are high-mass stellar systems whose spectral energy distribution
contains a significant and persistent non-thermal component, at energies above 1 MeV and up to the TeV domain. Only a handful of these
objects are currently known (Dubus 2013; Paredes et al. 2013). Among this scarce group, one finds a dominant presence
of luminous, emission line optical stars with Oe or Be spectral type. Their unseen compact companion can be
either a neutron star or a black hole. Here, we will broadly refer to these systems as Be/$\gamma$-ray binaries.
They can also be considered as a sub-class of the more numerous normal
%
Be/X-ray binaries, which contain more than 90 confirmed and suspected objects
(Reig 2011), but detected only up to keV energies.

At present there are three confirmed Be/$\gamma$-ray binaries so far -- LS~2883/PSR~B1259$-$63 (Aharonian et al. 2005), 
\lsi / 2CG~135+01 (Albert et al. 2009), and MWC~148/HESS~J0632+057 (Aharonian et al. 2007).
Some other binaries may be related to this group,  such as 
the system  MWC~656/ AGL~J2241+4454 with GeV transient $\gamma$-ray emission 
(Lucarelli et al. 2010; Casares et al. 2012).

The nature of the compact object is certainly known in  PSR~B1259$-$63 
(e.g. Chernyakova et al. 2015 and references therein). 
It is a neutron star acting as radio pulsar with a period of  47.76 ms (Shannon et al. 2014). 
The neutron star has a mass  $\sim 1.4$~$M_\odot$ and 
is orbiting a  O9.5Ve  star with mass  $M_1 \approx  30$~$M_\odot$
(Negueruela et al. 2011). 
For MWC~656, Casares et al. (2014) have analyzed  the radial velocities variability and 
found convincing evidences for a black hole of 3.8 to 6.9 $M_\odot$
orbiting an  B1.5-B2~IIIe primary   with $M_1 \approx  10-16$~$M_\odot$. 
More information on the compact object masses is needed to better
 discriminate among different possible theoretical scenarios for high energy emission.
 These mainly include  the magnetospheric pulsar model and the microquasar jet 
 model where a black hole could be expected (see e.g. Paredes et al. 2013).
 Other alternative scenarios could also play a role here, such as the
propeller regime in neutron stars proposed more
than three decades ago as a possible mechanism for TeV $\gamma$-ray emission  (Wang \& Robertson 1985)


In this work, we estimate the range of masses for the compact objects in two  Be/$\gamma$-ray binaries --
the systems  \lsi\ and MWC~148.

\section{Mass of the compact object}
\label{calc}

In the following, we will assume that the inclination of the orbit $i_{orb}$ is  approximately 
equal to the inclination of the Be star equatorial plane with respect to the line of sight within a few degrees. 
The question that it is plausible assumption  is shortly  addressed in Sect.\ref{D.1}.
In particular, we apply Kepler's third low:  
\begin{equation}
P_{orb}^2 =\frac{4 \pi ^2 (a_1+a_2)^3}{G(M_1 + M_2)}  \label{kepler}
\end{equation} 
where $G$ is the gravitational constant,
$P_{orb}$ is the orbital period of the binary, 
$a_1$ is the semi-major axis of the orbit of the primary,
$a_2$ is the semi-major axis of the orbit of the secondary,
$M_1$ is mass of the primary, 
and $M_2$ is the mass of the compact object.

For the primary star mass, we  use the best information available. 
In the \lsi\ case,  an appropriate range of values was derived from
its spectral type and the latest calibrations 
by Hohle et al. (2010)  based on revised Hipparcos data. 
In the MWC~148 case, the most accurate primary mass values come from
the comparison of its average optical spectrum with several grids of stellar models (Aragona et al. 2010).
 
Given a pair $(M_1, M_2)$ and the system orbital period, we compute the relative semi-major axis $a = a_1 + a_2$ using Eq. \ref{kepler}.
From published Doppler radial velocity observations, we also have an estimate of the projected semi-major axis $a_1 \sin{i_{orb}}$
of the optically visible primary star. This parameter can be de-projected using the assumed value of the orbital inclination.
Then, we can obtain the secondary semi-major axis as $a_2 = a - a_1$, and get an 
estimate of the secondary star mass as
$M_2 = M_1 (a_1/a_2)$. While keeping $M_1$ fixed, 
the procedure is iterated until the secondary mass converges within a dozen iterations.
The calculation is consequently repeated across the whole range of allowed primary mass values.

A consistency check of the procedure is that the resulting $M_2$ estimate has to be above the strict lower limit to the compact object
mass provided by the well known concept of mass function. 
This is given observationally by the following combination of projected semi-major axis and orbital inclination:

\begin{equation}
f(M_2)  = \frac{4 \pi^2 (a_1 \sin{i_{{orb}}})^3}{G P_{orb}^2}
\end{equation}


\subsection{\lsi}
\label{est.1}

\begin{figure}    
   \vspace{8.7cm}     
   \includegraphics{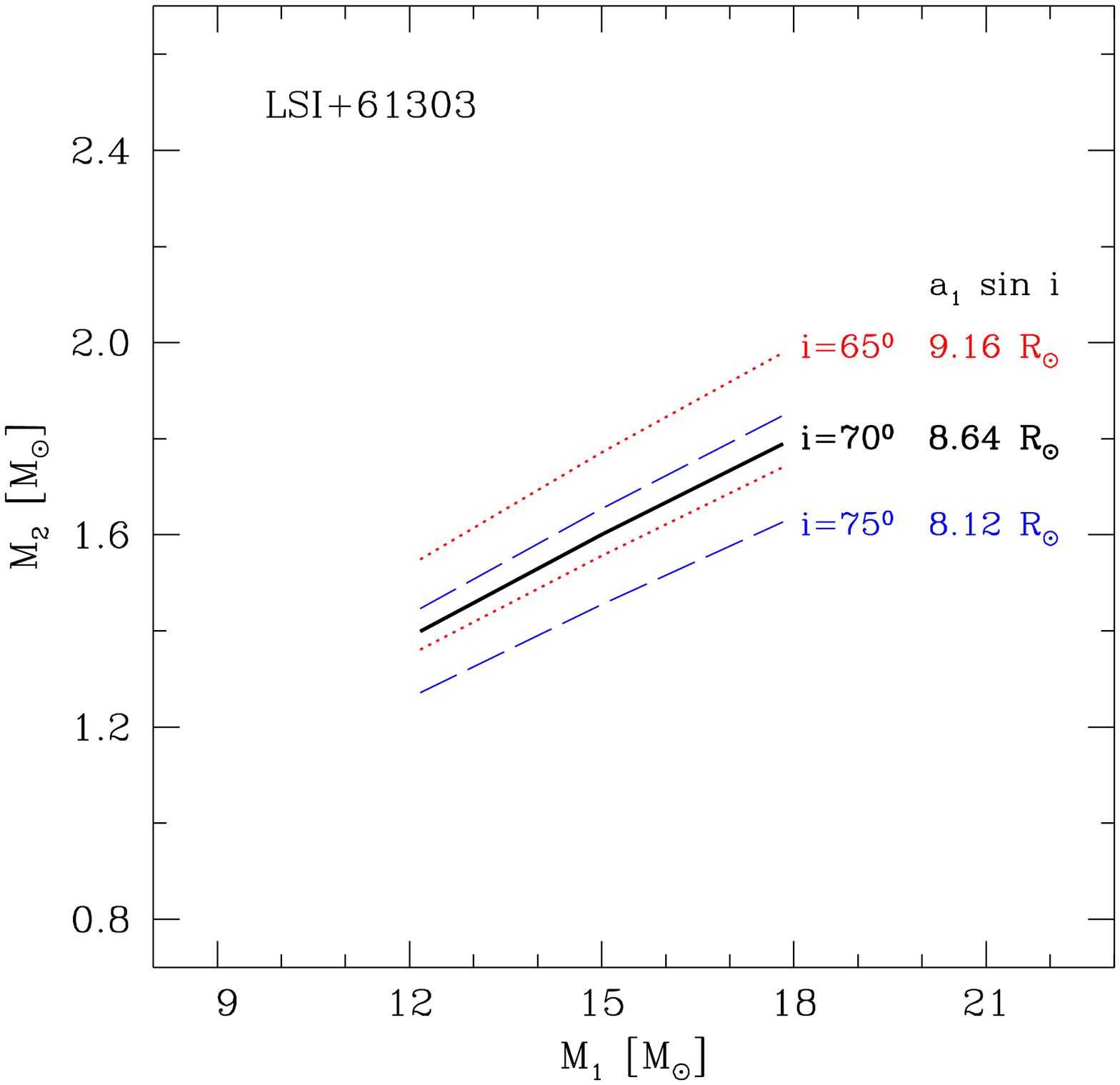}  
   \caption[]{Mass of the compact object versus the mass of the primary for the $\gamma$-ray
    binary \lsi.}
   \label{f1.lsi}      
   \vspace{8.7cm} 
   \includegraphics{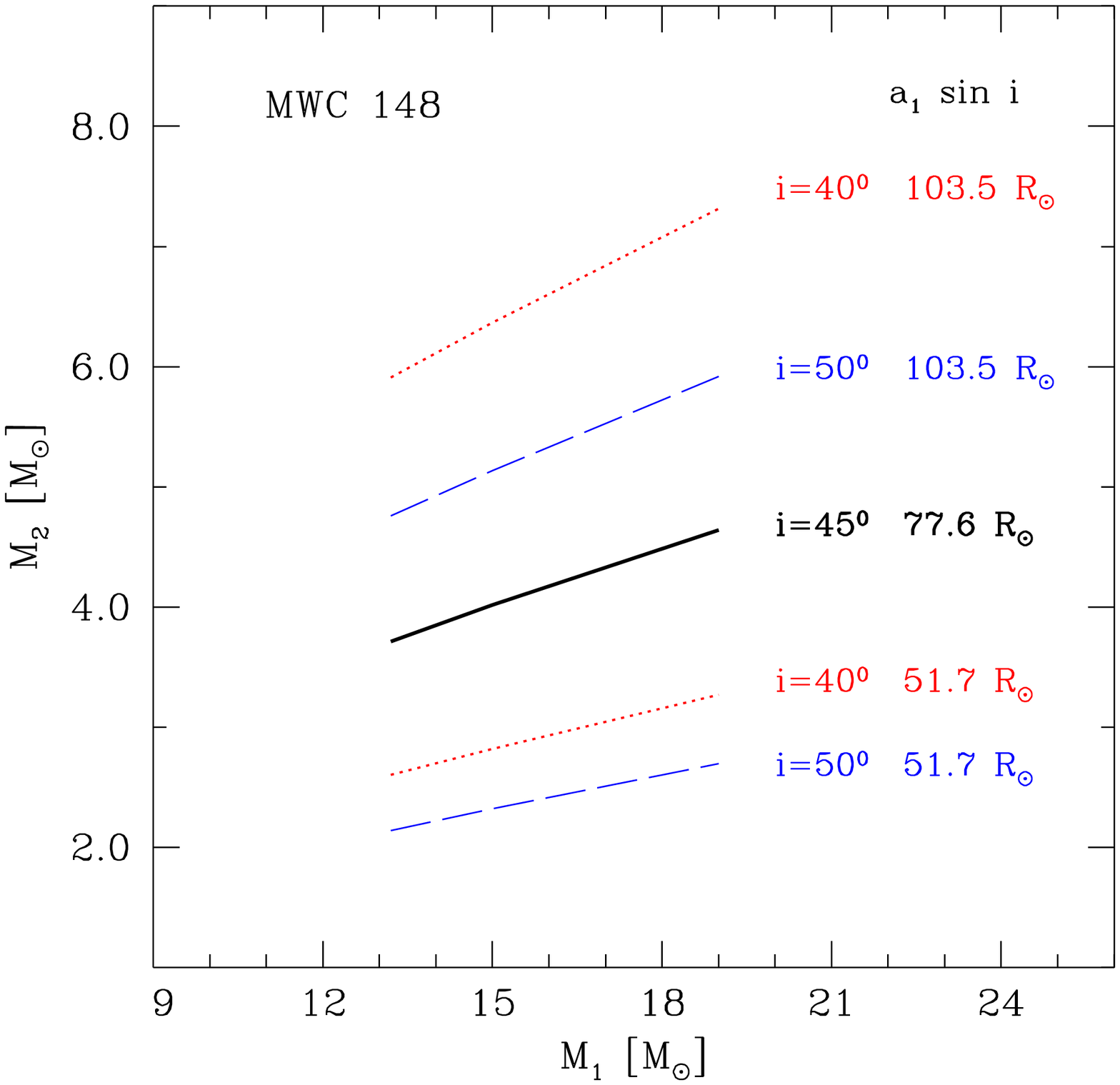}  
   \caption[]{Mass of the compact object versus the mass of the primary for the $\gamma$-ray
    binary MWC~148.}
   \label{f1.MWC148}      
 \end{figure}

From a radio survey of the galactic plane, 
\lsi\  (V615~Cas) was first proposed by Gregory \& Taylor (1978) 
as a $\gamma$-ray source in the $COS \; B$ satellite 
catalogue (Swanenburg et al. 1981). It became a confirmed TeV many years later (Albert et al. 2006).
A Bayesian analysis of radio observations gives the orbital period of the binary as $P_{orb}=26.4960 \pm 0.0028$~d (Gregory 2002). 
The orbital eccentricity is $e \simeq 0.537$, obtained on the basis of the radial velocity measurements 
of the primary  (Casares et al. 2005; Aragona et al. 2009).    
The inclination of the primary star Be disc in \lsi\ to the line of sight is probably   $ i_{Be} \sim  70^{\circ}$ according to Zamanov et al. (2013).
Aragona et al. (2009) give $a_1 \sin i_{orb} = 8.64 \pm 0.52$~\rsun. 
For the primary, Grundstrom et al. (2007) suggested a B0V star. 
A B0V star is expected to have on average $M_1 \approx 15.0 \pm 2.83 $~\msun\ (Hohle et al. 2010). 
We calculated $M_2$ as described above for a few sets of parameters $a_1 \sin i_{orb} = 8.12$, 8.64, 9.16~\rsun\ and
$i_{orb} =65^{\circ}$, $70^{\circ}$, $75^{\circ}$.  
The specified lines are plotted in Fig.~\ref{f1.lsi}. 
The red (dotted) lines are for  $a_1 \sin i_{orb} = 8.12$~\msun,  $i=65^0$
and   $a_1 \sin i_{orb} = 9.16$~\msun, $i=65^{\circ}$. 
The blue (dashed) lines are for  $i=75^0$.
The black (solid) line represents  $a_1 \sin i_{orb} = 8.64$~\rsun, $i_{orb} =70^0$, corresponding to the average values 
of separation and inclination. 

Assuming a B0V star with mass in the range $ 12.17$~\msun $ <  M_1 < 17.83 $~\msun, we estimate the mass of the compact object 
in the range $1.27 < M_2 < 1.98$~\msun, with most likely value $M_2 \approx 1.6$~\msun.  

\subsection{MWC~148}
\label{est.2}

MWC~148  (HD 259440)   was identified as the counterpart of the variable TeV source HESS J0632+057 (Aharonian et al. 2007, 
Maier \& for the VERITAS Collaboration 2015). 
We adopt  $P_{orb} = 315 ^{+6}_{-4}$~d derived from the X-ray data (Aliu et al. 2014), which is consistent with the 
previous result of $321 \pm 5$ days (Bongiorno et al.  2011). 
For this object,  Aragona et al. (2010) derived  $M_1 = 13.2 - 19.0$~\msun\ from their  spectral model fits.
From radial velocity measurements, 
 Casares et al. (2012) also estimated  $a_1 \sin i_{orb} = 77.6 \pm 25.9$~\rsun\ with an eccentricity of $e=0.83$.

The optical emission lines of MWC~148 are very similar to those of the bright well-known Be star $\gamma$~Cas (Zamanov et al. 2016).
All detected lines in the optical spectral range 4100 - 7500 \AA\ (Balmer lines, HeI lines and FeII lines) 
have similar  intensities, profiles, equivalent widths, and even a remarkable "wine-bottle" structure is  apparent
in the H$\alpha$ line profile. 
The emission lines are the most sensitive to the footpoint density and inclination angle (Hummel 1994). 
Based on such a strong resemblance, we consider that the Be star geometry in MWC~148 
should be similar to that of $\gamma$~Cas, 
for which the inclination is $43^{\circ} \pm 3^{\circ}$ (Poeckert \& Marlborough 1978;  Clarke 1990).
Therefore, we will proceed with our estimates  of the compact object mass
in MWC~148 by adopting different inclinations in the vicinity of this value.

We calculate $M_2$ for  different sets of parameters 
$a_1 \sin i_{orb} = 51.7$, 77.6, 103.5~\rsun\ and  $i_{orb} =40^{\circ}$, $45^{\circ}$, $50^{\circ}$.
The lines corresponding to these values are plotted in
Fig.~\ref{f1.MWC148}.
The red (dotted) lines are for $i=40^{\circ}$. 
The blue (dashed) lines are for $i=50^{\circ}$.
The black solid line represent  $a_1 \sin i_{orb} = 77.6$~\rsun, 
$i_{orb} =45^{\circ}$, corresponding to the average values of separation and inclination. 

Assuming  a mass of primary star  in the range $ 13.2 \le  M_1 \le 19.0 $~\msun, we estimate mass of the compact object 
in the range $2.1$~\msun$ < M_2 < 7.3$~\msun, with most likely value $M_2 \approx 4.0$~\msun.


\section{Discussion}
\label{D.1} 

The two $\gamma$-ray binaries discussed here have non-zero eccentricities and  misalignment 
between the spin axis of the primary component  and the spin axis of the binary orbit could be theoretically possible 
(Brandt \& Podsiadlowski 1995; 
Okazaki \& Hayasaki 2007; 
Martin et al. 2014). 
However, if a significant misalignment existed, then we would expect to see  considerable variability in the 
H$\alpha$ emission line at the time when the compact object crosses the circumstellar disc -- 
twice in  each orbital period. 
No such variability is detected in the observations of  H$\alpha$ emission, 
which means that any misalignment is less than the opening half-angle of the circumstellar disc. 
The opening half-angle of the  Be stars circumstellar disc are $\approx \! 10^{\circ}$ (Tycner et al. 2006; Cyr et al. 2015), 
and in Sect.~\ref{calc},  we have 
supposed that the orbital plane coincides with  the equatorial plane of the Be star within a few degrees. Therefore, our main assumption
in this work appears to be justified at least for the two systems being considered.
Our derived mass ranges are also dependent on additional 
assumptions on the physical properties of non-degenerate stars, specially the mass, according to the most recent data available.

Strict lower limits to the mass of the compact objects are set from the mass functions of the
different spectroscopic orbital solutions.  
For \lsi\, the   mass function  is  $f(M_2) =  0.0124 \pm 0.0022$~$M_\odot$ (Aragona et al. 2009), 
for MWC~148  it is  $f(M_2) = 0.06 ^{+0.15} _{-0.05}$  $M_\odot$ (Casares et al. 2012). 
The resulting mass ranges for both objects are, of course, safely above these values and therefore
consistent with what is know from radial velocity observations.

The masses of neutron stars ($M_{NS}$) measured in binary stars are in the range  0.9~\msun\  $< M_{NS} <2.7$~\msun\ 
({\"O}zel et al. 2012). The compact stars with a mass between 1.4~\msun\  (Chandrasekhar limit) and 2.8~\msun\ 
should be neutron stars (e.g. Chamel et al.  2013). 
The  mass ranges  calculated in Sect.\ref{calc}
point to the compact object in \lsi\  being most probably  a neutron star with a mass $\approx 1.6$~\msun. 
The spin period of the neutron star is expected to be $P_{spin} \approx 0.05 - 0.15$~s (Maraschi \& Treves 1981; 
Zamanov 1995), although the observational  search for pulsations
have not confirmed it yet 
(Coe et al. 1982;  Peracaula et al. 1997;  McSwain et al. 2011;  Ca{\~n}ellas et al. 2012).

There is a maximum  mass a neutron star may have (e.g.  Bombaci 1996).  
Antoniadis et al. (2016) considering the mass function of neutron stars and
mass measurements in binary millisecond pulsar establish that this maximum mass is of about 2.15~\msun.
The compact stars with a mass above the Tolman-Oppenheimer-Volkoff limit should be black holes. 
The measured masses of Galactic black holes are in the range  2.5-15~\msun\ ({\"O}zel et al. 2010). 
Our estimate of the mass of the compact object in MWC~148 (Sect.~\ref{est.2}) points to that 
it is likely to be a black hole with mass $\sim  4.0$~\msun. 
The calculated mass range for \lsi\ is  narrower than that of MWC~148, mainly because the projected semimajor axis 
$a_1 \sin i_{orb}$ is known with a considerably better accuracy, 
6\%  and 33\%  for \lsi\ and MWC~148, respectively. 

In most systems with  an early-type optical companion, $\gamma-$rays are usually believed to arise from 
the interaction between the stellar wind of the primary
and a pulsar magnetosphere instead of a black hole. However, an active debate is still open with both pulsar
and microquasar models in dispute (see e.g. Dubus 2013; Massi \& Torricelli-Giamponi 2016).
Both interpretations are competing to explain not only the $\gamma$-ray emission, but also the changing milli-arscecond radio structures observed
with interferometric techniques. If future multiwavelength
observations confirm the black hole nature of the companion star in MWC148 proposed here, this
would have important consequences on our general understanding of $\gamma$-ray binaries.
The $\gamma$-ray binary class could be then a multiface phenomenon with very different physical scenarios
coexisting in different systems.

\section{Conclusions} 
From the above considerations,   it appears that:
(i) the compact object in \lsi\  is most probably  a neutron star with  mass $\sim  1.6$~\msun, 
(ii) the compact object in MWC~148 is likely to be a black hole  with a mass $\sim  4.0$~\msun.  

The proposed non-uniform natures of the compact object in these two system  suggests that  different
physical scenarios, accounting for  very high energy emission in binary systems, can actually
take place in real systems.

%


\acknowledgements
This work was partly supported by grant AYA2013-47447-C3-3-P from the Spanish Ministerio 
de Econom\'{\i}a y Competitividad (MINECO), 
and by the Consejer\'{\i}a de Econom\'{\i}a, Innovaci\'on, 
Ciencia y Empleo of Junta de Andaluc\'{\i}a under research group FQM-322, 
as well as FEDER funds.


\begin{thebibliography}{}
\bibitem[Aharonian et al.(2006)]{2006A&A...460..743A} Aharonian, F., Akhperjanian, A.~G., Bazer-Bachi, A.~R., et al.\ 2006, \aap, 460, 743 
\bibitem[Aharonian et al.(2007)]{2007A&A...469L...1A} Aharonian, F.~A., Akhperjanian, A.~G., Bazer-Bachi, A.~R., et al.\ 2007, \aap, 469, L1 
\bibitem[Albert et al.(2006)]{2006Sci...312.1771A} Albert, J., Aliu, E., Anderhub, H., et al.\ 2006, Science, 312, 1771
\bibitem[Albert et al.(2009)]{2009ApJ...693..303A} Albert, J., Aliu, E., Anderhub, H., et al.\ 2009, \apj, 693, 303  
\bibitem[Antoniadis et al.(2016)]{2016arXiv160501665A} Antoniadis, J., Tauris, T.~M., Ozel, F., et al.\ 2016, \apj,  sumbitted,  arXiv:1605.01665 
\bibitem[Aragona et al.(2009)]{2009ApJ...698..514A} Aragona, C., McSwain, M.~V., Grundstrom, E.~D., et al.\ 2009, \apj, 698, 514 
\bibitem[Aragona et al.(2010)]{2010ApJ...724..306A} Aragona, C., McSwain, M.~V., \& De Becker, M.\ 2010, \apj, 724, 306 
\bibitem[Brandt \& Podsiadlowski(1995)]{1995MNRAS.274..461B} Brandt, N., \& Podsiadlowski, P.\ 1995, \mnras, 274, 461 
\bibitem[Ca{\~n}ellas et al.(2012)]{2012A&A...543A.122C} Ca{\~n}ellas, A., Joshi, B.~C., Paredes, J.~M., et al.\ 2012, \aap, 543, A122 
\bibitem[Bombaci(1996)]{1996A&A...305..871B} Bombaci, I.\ 1996, \aap, 305, 871
\bibitem[Casares et al.(2005)]{2005MNRAS.360.1105C} Casares, J., Ribas, I., Paredes, J.~M., Mart{\'{\i}}, J., \& Allende Prieto, C.\ 2005, \mnras, 360, 1105 
\bibitem[Casares et al.(2012)]{2012MNRAS.421.1103C} Casares, J., Rib{\'o}, M., Ribas, I., et al.\ 2012, \mnras, 421, 1103 
\bibitem[Casares et al.(2014)]{2014Natur.505..378C} Casares, J., Negueruela, I., Rib{\'o}, M., et al.\ 2014, \nat, 505, 378 
\bibitem[Chamel et al.(2013)]{2013IJMPE..2230018C} Chamel, N., Haensel, P., Zdunik, J.~L., \& Fantina, A.~F.\ 2013, 
          International Journal of Modern Physics E, 22, 1330018 
\bibitem[Chernyakova et al.(2015)]{2015MNRAS.454.1358C} Chernyakova, M., Neronov, A., van Soelen, B., et al.\ 2015, \mnras, 454, 1358 
\bibitem[Clarke(1990)]{1990A&A...227..151C} Clarke, D.\ 1990, \aap, 227, 151 
\bibitem[Coe et al.(1982)]{1982ApL....23...17C} Coe, M.~J., Bowring, S.~R., Hall, C.~J., \& Stephen, J.~B.\ 1982, \aplett, 23, 17 
\bibitem[Cyr et al.(2015)]{2015ApJ...799...33C} Cyr, R.~P., Jones, C.~E., \& Tycner, C.\ 2015, \apj, 799, 33 
\bibitem[Dubus(2013)]{2013A&ARv..21...64D} Dubus, G.\ 2013, \aapr, 21, 64 
\bibitem[Eger et al.(2016)]{2016MNRAS.457.1753E} Eger, P., Laffon, H., Bordas, P., et al.\ 2016, \mnras, 457, 1753 
\bibitem[Gregory \& Taylor(1978)]{1978Natur.272..704G} Gregory, P.~C., \& Taylor, A.~R.\ 1978 \nat, 272, 704
\bibitem[Gregory(2002)]{2002ApJ...575..427G} Gregory, P.~C.\ 2002, \apj, 575, 427 
\bibitem[H.~E.~S.~S.~Collaboration et al.(2015)]{2015A&A...577A.131H} H.~E.~S.~S.~Collaboration, Abramowski, A., Aharonian, F., et al.\ 2015, \aap, 577, A131
\bibitem[Hohle et al.(2010)]{2010AN....331..349H} Hohle, M.~M., Neuh{\"a}user, R., \& Schutz, B.~F.\ 2010, Astronomische Nachrichten, 331, 349 
\bibitem[Hummel(1994)]{1994A&A...289..458H} Hummel, W.\ 1994, \aap, 289, 458 
\bibitem[Maraschi \& Treves(1981)]{1981MNRAS.194P...1M} Maraschi, L., \& Treves, A.\ 1981, \mnras, 194, 1P 
\bibitem[Martin et al.(2014)]{2014ApJ...790L..34M} Martin, R.~G., Nixon, C., Armitage, P.~J., Lubow, S.~H., \& Price, D.~J.\ 2014, \apjl, 790, L34 
\bibitem[McSwain et al.(2011)]{2011ApJ...738..105M} McSwain, M.~V., Ray, P.~S., Ransom, S.~M., et al.\ 2011, \apj, 738, 105 
\bibitem[Maier \& for the VERITAS Collaboration(2015)]{2015arXiv150805489M} Maier, G., \& for the VERITAS Collaboration 2015, 34th International Cosmic Ray Conference 
(ICRC2015), arXiv: 1508.05489
\bibitem[Massi \& Torricelli-Giamponi(2016)]{2016A&A...585A.123M} Massi, M., Torricelli-Giamponi, C.\ 2016, \aap, 585, A123
\bibitem[Negueruela et al.(2011)]{2011ApJ...732L..11N} Negueruela, I., Rib{\'o}, M., Herrero, A., et al.\ 2011, \apjl, 732, L11 
\bibitem[Okazaki \& Hayasaki(2007)]{2007ASPC..361..395O} Okazaki, A.~T., \& Hayasaki, K.\ 2007, Active OB-Stars: Laboratories for Stellare and Circumstellar Physics, 361, 395 
\bibitem[{\"O}zel et al.(2010)]{2010ApJ...725.1918O} {\"O}zel, F., Psaltis, D., Narayan, R., \& McClintock, J.~E.\ 2010, \apj, 725, 1918 
\bibitem[{\"O}zel et al.(2012)]{2012ApJ...757...55O} {\"O}zel, F., Psaltis, D., Narayan, R., \& Santos Villarreal, A.\ 2012, \apj, 757, 55 
\bibitem[Paredes et al. (2013)]{2013APh....43..301P} Paredes, J. M., Bednarek, W., Bordas, P., et al. \ 2013, APh, 43, 301
\bibitem[Peracaula et al.(1997)]{1997A&A...328..283P} Peracaula, M., Mart\'{\i}, J., \& Paredes, J.~M.\ 1997, \aap, 328, 283 
\bibitem[Poeckert \& Marlborough(1978)]{1978ApJ...220..940P} Poeckert, R., \& Marlborough, J.~M.\ 1978, \apj, 220, 940 
\bibitem[Shannon et al.(2014)]{2014MNRAS.437.3255S} Shannon, R.~M., Johnston, S., \& Manchester, R.~N.\ 2014, \mnras, 437, 3255 
\bibitem[Swanenburg et al.(1981)]{1981ApJ...243L..69S} Swanenburg, B.~N., Bennett, K., Bignami, G.~F., et al.\ 1981, \apjl, 243, L69 
\bibitem[Tycner et al.(2006)]{2006AJ....131.2710T} Tycner, C., Gilbreath, G.~C., Zavala, R.~T., et al.\ 2006, \aj, 131, 2710 
\bibitem[Wang \& Robertson(1985)]{1985A&A...151..361W} Wang, Y.-M., \& Robertson, J.~A.\ 1985, \aap, 151, 361 
\bibitem[Zamanov(1995)]{1995MNRAS.272..308Z} Zamanov, R.~K.\ 1995, \mnras, 272, 308 
\bibitem[Zamanov et al.(2013)]{2013A&A...559A..87Z} Zamanov, R., Stoyanov, K., Mart{\'{\i}}, J., et al.\ 2013, \aap, 559, A87 
\bibitem[Zamanov et al.(2016)]{2016BlgAJ..24...40Z} Zamanov, R., Stoyanov, K., \& Mart{\'{\i}}, J.\ 2016, Bulgarian Astronomical Journal, 24, 40 (arXiv:1509.04191)

\end{thebibliography}
\end{document}